# AN INTERACTIVE EMPIRICAL APPROACH
# to the
# VALIDATION of SOFTWARE PACKAGE SPECIFICATIONS

S.D. Fraser and P.P. Silvester
CAD Lab, Department of Electrical Engineering,
McGill University, Montreal, Quebec, Canada

**Abstract**

The objective of this research is the development of a practical system to manipulate and validate software package specifications. The validation process developed is based on consistency checks. Furthermore, by means of scenarios, the customer will be able to interactively experience the specified system prior to its implementation. Functions, data, and data types constitute the framework of our validation system. The specification of the Graphical Kernel System (GKS) is a typical example of the target software package specifications to be manipulated.

## 1. INTRODUCTION

Problem definition or specification is a complex task in the construction of any structure. If a problem cannot be described, then a 'solution' is impossible to obtain. However, it is acknowledged that the existence of a specification does not necessarily imply that the solution sought can be obtained. Our objective is to create a specification framework that is practical and requires little or no formal mathematical preparation for its use. This does not imply that a mathematical basis is lacking from the system, merely that it is 'hidden' from the customer.

In software engineering a specification may be considered as a document that describes the "WHAT" of the system to be developed. Many authors [1,2,3,6,7,8,9] describe the desired qualities in a specification. In general, they agree that it should be as vague as possible to permit maximum freedom to the implementor. A software package is a collection of subroutines and their associated data structure that together create a specific application environment, e.g., a computer graphics interface. Specification validation is the process that determines if the proposed specification is a consistent and accurate reflection of customer desires.

This research was conducted in an environment where several moderately sized software developments are underway. While the systems developed are not necessarily complex, their size introduces substantial 'book-keeping' problems. For example, a typical software package system developed would be a finite element, interactive, graphic based system.

Our specification validation system, SVSP (for specification validation of software packages), is not intended as an answer to all specification problems. It should be viewed as a tool to facilitate the specification of software packages and improve communication between the specifier and the customer. SVSP is essentially a platform to organize functionality and data element interaction at the concept (specification) level.

If SVSP is able to inform the specifier that a particular data element referenced by a certain function is non-existent or improperly manipulated, or that the representation of the specification in scenario form is not quite what the customer had in mind, then it will have served a practical purpose. To some extent this is an empirical approach because it is difficult to precisely determine a basis for customer acceptance/rejection [5,10].

SVSP addresses this problem by performing consistency checks between functions and data elements. These checks involve the determination of existence and the uniformity of references. Consistency with user needs is determined both statically and dynamically by means of scenarios.

The user is able to manipulate function and data element descriptions methodically. These manipulations are similar to those associated with elementary set theory. A software package specification is viewed as a collection of functions, data elements, and data types. In a specification, functions and data elements do not necessarily have a direct relationship with the subroutines and variables of implemented computer code. Validation takes place on two different levels: firstly as a determination of the internal consistency of the proposed specification, and then a check with the customer to determine their acceptance.

To determine if our proposed validation system is practical, the ISO specification of GKS was examined. GKS contains approximately two hundred functions, one thousand data elements, and other associated information. GKS is a lengthy specification (266 pages [4]). It also imposes real restrictions, i.e., we have assumed that GKS typifies a reasonably well written specification using only the English language as opposed to a specification written in a formal language.

We note that our approach, while based on set theory, it may not be as elegant as other methods. To some extent this is due to the intangible relationship between a customer and a specification. This is where empirical relationships form an important part of our methodology.

**2. FUNCTION STRUCTURE**

The objective of a practical system is action. Action is typified by the representation of a relation between initial and final conditions. The initial and final conditions may be described by data elements and their associated data types. Descriptions of actions (functions) and data must be detailed to a limited extent.

A function is considered as a function identifier, a set of parameter identifiers, a set of function effect identifiers, and a function classification. The unique function identifier is used to reference the function in all operations by SVSP. The parameter list is the set of data elements that are used as either input or output to the function. This includes data elements explicitly included in the argument list of a function and those implicitly affected by the function effects.

The function effects are descriptions of operations performed on data elements by a function. They describe either an explicit transformation between input and output data elements, or an 'abstract' operation. The required status of the data elements referenced, both before and after the effect has been performed forms the basis of an 'effect.' The actual values of data elements are not required, neither is a description of how a calculation is to be performed. That description may be left to an implementor. For example, the square root function may be calculated by many different algorithms.

The function classification is a set of four descriptors that are used in the manipulation process of the specification. For example, functions may be grouped by type of operation, or by the state in which they may be performed.

## 3. DATA ELEMENT STRUCTURE

Data elements are descriptions of the information to be manipulated by the system specified. A data element is considered to consist of: a data element identifier: a data type identifier: data value restrictions: and a value. The identifiers are used for reference purposes. The restrictions take two forms: there may be restrictions on the actual value that a data element may have as given in the specification. Similarly, there are always restrictions on a data element's status. A data element's status depends on its allocation, definition, and value.

If a data element is allocated, it may be manipulated. If a data element has been assigned a value (which may not actually be known), it is considered defined. If the actual value is known, further manipulations are permitted, e.g., comparisons, operations. It is noted that these restrictions will flag data elements that are used as input, when they are neither allocated nor defined. The restrictions on the actual value of a parameter may be in the form of range restrictions for real numbers or integers, and length for character strings.

A combination of the restrictions noted above is also embedded in the effect/parameter descriptions associated with functions. For example, a function might require that a particular data element falls within a certain range, e.g., non-negative. This restriction should be checked for consistency, as a subset of the possible restrictions maintained at the data element description level. Restrictions on integers and real numbers are described in terms of inequalities.

While SVSP offers flexibility in the definition of functions, and data elements as described above, at present only a small number of data types is supported. These include: real numbers, integers, and character strings. More complex data types may be constructed from these simple types. Similarly, only a limited set of operations may be performed on data elements; this is a reflection of the need to describe only the 'what' of the function in contrast to the actual 'how.'

## 4. COMPONENTS OF SVSP

SVSP consists of three principal components, an editor, a query display, and a scenario generator. The editor adds, modifies, or deletes information within the specification structure. This includes both function and data-related descriptions. The editor contains a checker which is the comparator of the system. It checks for the uniqueness of definition, the existence of references, and the agreement of restrictions. The scenario generator offers the user an opportunity to experience the system specified with the simulations of individual function effects. The query display offers direct access to information on the specification, for example, selective combinations of information on functions, data and/or data types.

The editor works in a piecewise fashion. When a particular definition is complete, a check must be performed on the proposed change before modification of the previously described specification by SVSP can occur. In this manner contradictions can be rationalized on an individual basis.

The checker is the comparator of the system and is completely transparent to the user of SVSP, i.e., it is a component of the editor. It checks for uniqueness of definition: existence of references:

and the consistency of restrictions. Each function/data-element is added/modified on an individual basis. This permits the system to be maintained as 'consistent' on a piecewise basis.

The 'experience' that the scenario generator offers the user enables specification changes to be made without the cost associated with implementation changes. The scenario generator permits a limited simulation and dynamic test of system functions, e.g., call sequences, data compatibility. Specifically, this scenario generator will offer a simulation of the function calls and their related effects. We reiterate that we are dealing with the specification of software packages, our concerns are therefore directed more to function-data relationships than to the problems in control encountered at the level of computer code.

Once the scenario generator is initiated, the user chooses a function to simulate in the specification under consideration. If the user is not aware of the functions available, or their required parameters, further information may be obtained from the query display. The functionality of the specified system may be thus evaluated from this simulation.

Information on the specification should be obtainable directly by the specifier at any stage of the system development. The information to be displayed may include: functions: data elements: data types: or selective combinations of functions, data, and/or data types.

## 5. THE IMPLEMENTATION OF SVSP

SVSP has been developed on a VAX 750 UNIX based system. While programming languages such as PROLOG or LISP were considered, they were not practical in our research environment. At present SVSP is implemented in C with the INGRES relational database and the CURSES full screen display support package. Tests using the specification of GKS have begun.

## 6. SUMMARY

We believe that SVSP will fill a gap between informal design methods and the formal specification languages. Each of these methodologies has a role to play and a specific target use. SVSP can be used without imposing an overly restrictive environment on the customer. It is a compromise between the informality of natural language and the formality of predicate calculus or other formal methods.

The system will serve a useful purpose. Any flaws in a system specification that are detected prior to implementation will save both time and money in development and maintenance costs. Although the literature describes many software development tools, most of them are aimed at verifying code developed from a specification, rather than validating a specification in relation to customer needs. Some work has been done, however these advances appear to be oriented mainly towards formal specification methods.

Many problems related to specification ambiguity or inconsistency should be alleviated simply by the application of SVSP's editor. This would be done without the expense of the application of apparent mathematical formalism. The application of elementary set theory to the manipulation of specifications appears to offer a reasonable compromise between formality and informality.

The authors wish to acknowledge the financial support of the Natural Sciences and Engineering Research Council of Canada.